# CURRENT STATUS OF THE LBNE NEUTRINO BEAM *

Craig Damon Moore, Ken Bourkland, Cory Francis Crowley, Patrick Hurh, James Hylen, Byron Lundberg, Alberto Marchionni, Mike McGee, Nikolai V. Mokhov, Vaia Papadimitriou, Rob Plunkett, Sarah Diane Reitzner, Andrew M Stefanik, Gueorgui Velev, Karlton Williams, Robert Miles Zwaska, Fermilab, Batavia, Il, USA

*Abstract*

The Long Baseline Neutrino Experiment (LBNE) will utilize a neutrino beamline facility located at Fermilab. The facility is designed to aim a beam of neutrinos toward a detector placed in South Dakota. The neutrinos are produced in a three-step process. First, protons from the Main Injector hit a solid target and produce mesons. Then, the charged mesons are focused by a set of focusing horns into the decay pipe, towards the far detector. Finally, the mesons that enter the decay pipe decay into neutrinos. The parameters of the facility were determined by an amalgam of the physics goals, the Monte Carlo modeling of the facility, and the experience gained by operating the NuMI facility at Fermilab. The initial beam power is expected to be ~700 kW, however some of the parameters were chosen to be able to deal with a beam power of 2.3 MW. The LBNE Neutrino Beam has made significant changes to the initial design through consideration of numerous Value Engineering proposals and the current design is described.

## INTRODUCTION

The LBNE neutrino beam needs to provide a wide band beam to cover the first and second neutrino oscillation maxima (.5 GeV to 4 GeV) and modeling has been used in the choice of system parameters to optimize the operation of the facility in this region. The initial operation of the system will be at a beam power incident on the production target of 700 kW however some of the initial implementation will have to be done in such a manner that operation at 2.3 MW can be achieved without retrofitting. The cooling systems for the chase, decay tunnel, and absorber must be designed so that an upgrade for 2.3 MW operations is achievable without cost implications other than increasing the capacity. The relevant radiological concerns: prompt dose, residual dose, air activation, tritium production have been extensively modeled and these issues have been used in the system design. An initial design of the full scope LBNE was successfully reviewed in March 2012 however shortly thereafter the Department of Energy (DOE) stated that the Project must be constructed in a staged manner to reduce the initial cost. A "Reconfiguration" process utilizing many Value Engineering proposals then led to a successful set of DOE reviews in October and November 2012 of the Phase I LBNE Project, and Critical Decision 1 (CD-1) was approved on December 10, 2012. This paper reviews the status of the Neutrino Beam part of the Phase I LBNE Project after the Reconfiguration.

## MODELING

A software model for the entire beamline has been coded into the MARS15 simulation package []. This model includes all the essential components of the target chase, decay channel, hadron absorber and the steel and concrete shielding in the present design. The MARS simulations are used specifically in the calculation of: (1) beam-induced energy deposition in components for engineering design, (2) prompt dose rates in halls and outside shielding, (3) residual dose rates from activated components, (4) radionuclide production in components, shield and rock, (5) horn focusing design and optimization of neutrino flux. Fig. 1 shows the unified MARS model of the Neutrino Beam with detailed views of the target chase and Absorber region. Calculated dynamic heat loads in the hottest components are 14 to 30 kW, with residual dose rates on contact ranging from 145 to 345 Rem/hr after 30-day irradiation and 1-day cooling.

A most important aspect of modeling at the present design stage is the determination of necessary shield thickness and composition. For example, ground water protection is a crucial consideration for the 204 m-long decay tunnel. Results from modeling calculations are essential in the design of all of the subsystems discussed in the following sections.

The neutrino beamline is a set of components and enclosures designed to efficiently convert the primary proton beam (up to 120 GeV per proton) to a neutrino beam (maximum flux at neutrino energies 2-3 GeV) aimed at the far detectors, 1280 km away. In order of placement, this design includes (1) a baffle to protect downstream components from mis-steered beam, (2) a target, (3) toroidal focusing horns for the secondary pions, and surrounding shielding, (4) a 204-m long, 4-m diameter decay pipe to allow pions to decay to neutrinos, and(5) an absorber at the end of the decay pipe to deal with uninteracted protons and residual pions. The level of detail incorporated into the present simulations is conveyed in Fig. 1.

___________________________________
*Work supported by the Fermilab Research Alliance, under contract DE-AC02-07CH11359 with the U.S. Dept of Energy.
! cmoore@fnal.gov

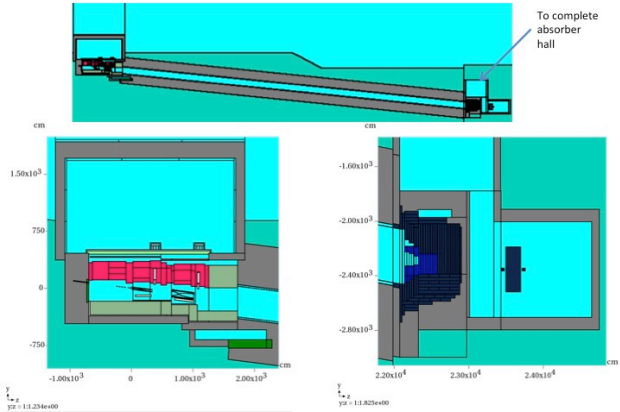

Figure 1. Unified MARS model of the Neutrino Beam with detailed views of the target chase and Absorber region.

## BEAM WINDOW, BAFFLE, AND TARGET DESIGN

The conceptual primary beam window design considers a 50 mm diameter, 0.2 mm thick partial hemispherical beryllium window, water-cooled in the 2.3 MW (1.5 mm spot size) case. A stainless steel, 117.5 mm O.D. Conflat flange with knife-edge seal provides a bolted connection to the primary beam pipe.

The baseline baffle design considers 700 kW beam energy with a view towards the beam power upgrade to 2.3 MW. The baseline baffle consists of ten 57 mm O.D. x 13 mm I.D. x 150 mm long graphite R7650 grade cores which are enclosed by a 150 cm long aluminum tube after annealing. The graphite baffle prevents mis-steered primary proton beam from causing damage to the horn neck and target cooling/support components, and the outer radius portion of the absorber. It must withstand the full intensity of the beam for a few pulses during the time needed to detect the mis-steered beam and terminate beam.

The conceptual target design for LBNE is based on experience with the Fermilab NuMI neutrino beam and studies for higher power beams. Graphite was used in the NuMI beam, and its performance was basically successful. However, over time five targets have failed due to ancillary components, and one has failed from ~10% graphite degradation. Graphite has been adopted as the baseline target material, but alternatives are under study.

The target uses a graphite core evolved from the NuMI designs and studies for a higher-power beam, developed at IHEP-Protvnio. The graphite core is segmented into short fins, 15.0 mm high, 6.4 mm wide and 20 mm in length. The total graphite length is 95 cm. Titanium water-cooling tubes are attached at the top and bottom of the fins, surrounded by a 3 cm diameter helium containment tube that has beryllium windows at the upstream and downstream ends.

## HORNS AND HORN POWER SUPPLY

The focusing of charged pions is produced by the toroidal magnetic field present in the volume between the co-axial inner and outer conductors of the horns with a pulsed current. The inner conductors of Horn 1 and Horn 2 have a double-paraboloid shape as depicted in Fig. 2.

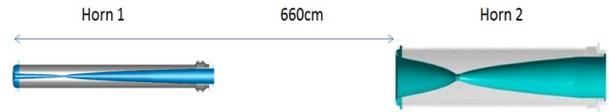

Figure 2. LBNE Horns

Horn conductors must withstand repetitive thermal and magnetic loading over tens of millions of beam/current pulses. The conductors are made of aluminum 6061-T6 and will be cooled by spray water. The interior of the horn contains an inert Argon atmosphere that is continuously purged to remove hydrogen and oxygen from dissociated water. Thermal and structural finite element analyses were carried out to guide the design and study the fatigue strength of the inner conductors. Horns will be supported and positioned by support modules, which are capable of aligning and servicing the horn by remote control.

The current NuMI horn power supply (Fig. 3) will be used for the LBNE horns.

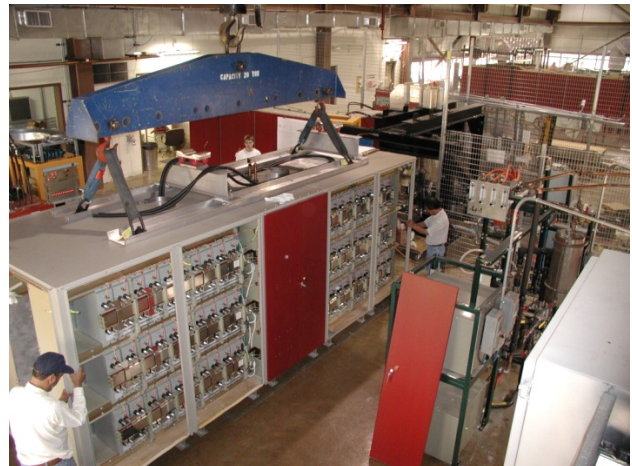

Figure 3. Completed NuMI horn power supply.

## TARGET SHIELD PILE

The shield surrounds the baffle, target and horns and leads directly to the decay pipe. The shield's purpose is to provide radiological protection for groundwater concerns, electronic component lifetime issues, and personal access issues with respect to residual radiation. It is composed of an inner layer of steel surrounded by concrete.

## DECAY PIPE AND HADRON ABSORBER

The decay pipe is the region where the mesons generated from the target decay into neutrinos. The dimensions (4-m diameter and 204-m length) of the pipe have been chosen based upon considerations of cost and maximizing of the neutrino flux in the desired energy range.

At the end of the decay pipe, the s - primary protons, non-decayed secondary hadrons (mostly $\pi$ and K-mesons) -must be absorbed to prevent them from entering the surrounding rock of the excavation and inducing radioactivity. The absorber structure is a pile of aluminum (Al), steel and concrete blocks, some of them water-cooled, MARS s 26% of the total beam power system The absorber design is further described in []. A fraction of the muons that are produced together with neutrinos from meson decays penetrate through the absorber, where they can be detected to monitor beam performance.

## RAW WATER SYSTEMS

Radioactive Water (RAW) systems are small volume systems that are used to take heat away from regions where much thermal heat load is developed. The systems must be robust and have containment capability for the entire capacity of the system.

General construction will be to ANSI Code B31.1 Process Piping, with rigorous weld inspection and radiography. Reservoir tanks will be specified as coded pressure vessels. Redundancy and containment will be used where deemed appropriate. Significant filtration and deionizing bottles will be used to minimize the particulate build-up in the water. RAW capture and makeup systems will be integrated into the design, so that down time and worker exposure for routine service may be minimized.

Intermediate systems will transfer heat from the RAW systems, to chilled water systems that take the heat to the surface for removal via chillers and/or cooling ponds. The intermediate systems also provide a second layer of containment for radioactivated water, in case of a breach of the primary heat exchangers.

## REMOTE HANDLING EQUIPMENT

Technical components installed in the Target Chase and in the Absorber Hall areas are subjected to intense radiation from the primary or secondary beam. The level of irradiation in some LBNE environments will reach levels that are unprecedented at Fermilab. Components to be handled, serviced, and/or stored range in size (from 0.25 $m^3$ to 25 $m^3$), weight (from 10 kg to 30,000 kg), and estimated dose rate (from 5 R/hr to 8000 R/hr on contact). Therefore remotely operated removal and handling systems are an integral part of the Target Hall and Absorber Hall designs. Since the remote handling systems are integrated into the infrastructure and cannot be upgraded after irradiating the Target Hall and Absorber Hall areas, they must be designed to be sufficient for 2.3 MW beam power.

Remote Handling Systems will include a remotely operated crane with redundant drives and a work cell enabling target and horn replacement in the Target Hall area. In addition, the neighboring service building will include a remote handling procedures mock-up area and a short-term radioactive component storage area.

## TRITIUM MITIGATION

Based upon experience with NuMI a system has been designed that will mitigate the expected production of tritium in such a manner that no detectable amount of tritium will be in the waters leaving the site boundaries. This is achieved through a combination of shielding and the use of barriers to prevent water moving between the beamline complex and the environment. As water is a major vector in which tritium can be transmitted, it is imperative that any water inside the beamline complex is kept separate from the environment. Furthermore, outside water should not be allowed to enter in order to maintain a low air humidity, which in turn minimizes the amount of airborne tritium.

The barrier system consists of multiple layers of water impermeable barriers interspaced with drainage systems. These will be deployed outside the critical areas of shielding. Having multiple impermeable barriers prevents water form entering if one of the barriers is breached. The drainage layers will collect any water that does manage to enter the barrier system.

## SUMMARY

A major redesign of the Neutrino Beam has been accomplished for LBNE Phase I, which provides a proper wide band neutrino, beam at the far detector with a significantly reduced cost while keeping the ability to upgrade to 2.3 MW.